\documentclass[12pt]{article}
\usepackage{amsmath} 
\usepackage{amssymb}
\usepackage{hhline}
\usepackage[scale={.75,.75}]{geometry}
\usepackage{url}
\usepackage{lscape}
\usepackage{caption}
\usepackage[numbers,sort&compress]{natbib}
\usepackage{epsfig}
\usepackage{multirow} 
\usepackage{color}
\usepackage{verbatim}
\usepackage{nicefrac}
\usepackage{upgreek}
\usepackage{bbm}
\usepackage{setspace}
\usepackage{pstricks}
\usepackage{psfrag}
\usepackage{color}

\usepackage{hyperref}% add hypertext capabilities

\renewcommand{\k}{\textbf{k}}

\newcommand{\V}{\mathcal{V}}

\newcommand{\Rrad}{{\rm R}_{\rm rad}}
\newcommand{\zend}{z_{\rm end}}
\newcommand{\zreh}{z_{\rm reh}}
\newcommand{\areh}{a_{\rm reh}}
\newcommand{\aend}{a_{\rm end}}
\newcommand{\rhoend}{\rho_{\rm end}}
\newcommand{\rhoreh}{\rho_{\rm reh}}
\newcommand{\rhotildegamma}{\tilde{\rho}_\gamma}
\newcommand{\wrehbar}{\bar{w}_{\rm reh}}
\newcommand{\wreh}{w_{\rm reh}}
\definecolor{green}{rgb}{0,0.5,0}

\begin{document}
\date{}
%\preprint{TUM-HEP-???/14}

\title{\vspace{-1.8cm} 
%{\normalsize TTK-12-04\hfill\mbox{}\\}
%\vspace{0.5cm}
\begin{flushright}
\vspace{-1cm}
{\scriptsize \tt TUM-HEP-1021/15, CAS-KITPC/ITP-455}  % 11-18
\end{flushright}
{\bf What can the CMB tell about the microphysics of cosmic reheating?
}
}

\author{
Marco Drewes\\
\footnotesize{Physik Department T70, Technische Universit\"at M\"unchen,}\\
\footnotesize{ James Franck Stra\ss e 1, D-85748 Garching, Germany}
}
\vspace{-0.5cm}
\maketitle
\vspace{-0.5cm}
\begin{abstract}
  \noindent In inflationary cosmology, cosmic reheating after inflation sets the initial conditions for the hot big bang. We investigate how CMB data can be used to study the effective potential and couplings of the inflaton during reheating and constrain the underlying microphysics. If there is a phase of preheating that is driven by a parametric resonance or other instability, then the thermal history and expansion history during the reheating era depend on a large number of microphysical parameters in a complicated way. In this case the connection between CMB observables and microphysical parameters can only established with intense numerical studies. Such studies can help to improve CMB constraints on the effective inflaton potential in specific models, but parameter degeneracies usually make it impossible to extract meaningful best-fit values for individual microphysical parameters.
If, on the other hand, reheating is driven by perturbative processes, then it can be possible to constrain the inflaton couplings and the reheating temperature from CMB data.
This provides an indirect probe of fundamental microphysical parameters that most likely can never be measured directly in the laboratory, but have an immense impact on the evolution of the cosmos by setting the stage for the hot big bang.  
\end{abstract}
%\newpage
\vspace{-0.5cm}
\begin{footnotesize}
\tableofcontents
\end{footnotesize}

%\pacs{xxx}

\section{Introduction}\label{Nepenthe}
The cosmic microwave background (CMB) is an extraordinary rich source of information about the early universe. Amongst others, the power spectrum of its temperature fluctuations have confirmed several predictions of cosmic inflation \cite{Starobinsky:1980te,Guth:1980zm,Linde:1981mu}. Though not being a strict proof, this makes the idea that the observable universe underwent a phase of accelerated cosmic expansion during its very early history very appealing. It is, however, to date entirely unknown what mechanism drove this rapid expansion. This is not only unsatisfactory for cosmologists, but also for particle physicists, who would like to see this mechanism embedded into a more general theory of nature.  Hence, there is strong motivation for cosmologists and particle physicists to look for information about inflation in the CMB.

Due to its simplicity, the probably most popular idea to explain inflation is that a negative equation of state was realised when the potential energy of the expectation value $\varphi\equiv\langle\phi\rangle$ of a scalar \emph{inflaton} field $\phi$ with a flat effective potential $\V(\varphi)$ dominated the energy density of the universe.
There exits many possible realisations of this idea, see e.g.\ \cite{Martin:2013tda} for a partial overview.\footnote{Meanwhile one should bear in mind that 
the mechanism behind cosmic inflation may lie far beyond any of the ideas presented in the literature so far, and even beyond our understanding of fundamental physics.}
For simplicity and illustrative purpose, we in the following focus on such models. The generalisation to models involving several fields is straightforward. 
The data from the Planck  satellite allows to impose considerable constraints on $\V(\varphi)$ \cite{Planck:2013jfk}. 
This certainly comprises one of the great successes of modern cosmology. 
However, the part of the potential $\V(\varphi)$ that can be probed with the CMB is relatively narrow.
Far away from this region, $\V(\varphi)$ may look very different from the function that provides the best fit to CMB data. 

Recently it has been pointed out that the CMB may also be used to constrain $\V(\varphi)$ near its minimum, i.e. in an entirely different region, via the dynamics during \emph{cosmic reheating} \cite{Cicoli:2012cy,Martin:2014nya,Dai:2014jja}. Cosmic reheating is the process in which $\varphi$ dissipates its energy density $\rho_\varphi$ into other degrees of freedom, usually via particle production during oscillations around its potential minimum  \cite{Traschen:1990sw,Shtanov:1994ce,Kofman:1994rk,Kofman:1997yn,Boyanovsky:1994me}. These are usually thought to be relativistic and dubbed "radiation" regardless of their spin, mass or other quantum numbers. 
Reheating  populates the universe with hot radiation and sets the initial conditions for the ``hot big bang'', i.e. the beginning of the radiation dominated era in cosmic history. Since the transfer of energy from $\varphi$ to other degrees of freedom necessarily involves interactions, one may also hope to learn something the inflaton couplings from reheating.

The reheating process takes a finite amount of time, it should be regarded as a separate \emph{reheating era} in cosmic history that begins when the universe stops accelerating (the equation of state exceeds $w>-1/3$) and ends when the energy density $\rho_\gamma$ of radiation exceeds $\rho_\varphi$. 
The end of the reheating era is not well-defined; common choices are the moment when $\rho_\gamma=\rho_\varphi$ or moment when the equation of state of the combined system of $\varphi$ plus radiation approaches $w=1/3-\delta$, with some arbitrarily chosen small parameter $\delta$. 
The information that could be obtained from this era is very rich. It is not only sensitive to the shape of $\V(\varphi)$ near its minimum, but also to the inflaton's coupling to matter and radiation.
The reheating process affects cosmology in at least three different ways:
\begin{itemize}
\item It affects the \emph{expansion history} because the equation of state during reheating is different from in inflationary or radiation dominated era.
\item It affects the \emph{thermal history} because it determines the temperature at the onset of the radiation dominated era, and there is usually a complicated (often far from equilibrium) thermal history during reheating itself. 
\item It may produce relics, such as gravitational waves, magnetic fields or topological defects. Even the baryon asymmetry of the universe or Dark Matter can be produced during reheating.
\end{itemize}
In this paper we use simple scenarios that have been discussed in the literature to examine how much information about reheating can be extracted from the CMB, 
focusing on the approach proposed in \cite{Martin:2010kz}. 
We also discuss how much one can learn about the underlying microphysics from this information.
The are different ways to pose these questions, for instance
\begin{itemize}
\item[1)] If one specifies a cosmological model  of inflation, which role does the reheating era play when extracting information about $\V(\varphi)$ during inflation from the CMB?
\item[2)] If one specifies a cosmological model of inflation, what can the CMB tell about the reheating phase?
\item[3)]  How can the imprint of reheating in the CMB help to determine which of the known models (if any) is realised in nature?
\item[4)]  If one specifies a particle physics model, can the information on the reheating phase that the CMB contains be used to improve constraints on model parameters?
\end{itemize}

\section{Setup and definitions}

\subsection{Some definitions}
Before addressing the questions 1)-4) in more detail, we need to define some terminology.
First we recall that during inflation, in very good approximation all energy is stored in the expectation value (or "one point function") $\varphi\equiv\langle\phi\rangle$ of the zero mode of the scalar inflaton field $\phi$.
Here the expectation value $\langle\ldots\rangle$ includes quantum as well as thermodynamic fluctuations, which can be important once the hot primordial plasma forms during reheating.   
\begin{itemize}
\item By a cosmological \emph{model of inflation} we mean the specification of the effective potential $\V(\varphi)$ that $\varphi$ feels during inflation and reheating.
The effective potential $\V(\varphi)$ usually has a different shape than the bare potential $V(\phi)$ that appears in particle physics models.\footnote{As usual, we include the mass term $\frac{1}{2}m_\phi^2\phi^2$ into $V(\phi)$.}
The bare potential defines a particle physicist's model of inflation $\mathcal{L}_{\rm infl}=\frac{1}{2}\partial_\mu\phi\partial^\mu\phi-V(\phi)$, but is not directly probed by cosmological observations. The quantity that enters the effective equation of motion (\ref{EOM}) for $\varphi$ in the early universe is the effective potential $\V(\varphi)$, which includes radiative and thermal corrections
\item By \emph{inflaton Lagrangian} $\mathcal{L}_\phi=\mathcal{L}_{\rm infl}+\mathcal{L}_{\rm int}$ we mean all terms in the Lagrangian of a particle physics model that contain $\phi$, including $\mathcal{L}_{\rm infl}$ and the inflaton's coupling to other degrees of freedom in $\mathcal{L}_{\rm int}$. We  characterise the strength of these interactions by a set $\lambda=\{\lambda_i\}$ of dimensionless parameters (e.g. effective coupling constants).
We collectively refer to the set of all degrees of freedom other than $\varphi$ as  $\mathcal{X}=\{\mathcal{X}_i\}$, without specifying the spin or other quantum numbers of the individual fields $\mathcal{X}_i$ that form the primordial plasma.
The set $\mathcal{X}$ includes all fields of the Standard Model (SM) of particle physics and possibly other fields (at least a Dark Matter candidate and an explanation for neutrino masses).
The $\mathcal{X}_i$ can have various types of interactions amongst each other, the strength of which we characterise by another set of dimensionless parameters $\alpha=\{\alpha_i\}$.
\item By \emph{particle physics model} $\mathcal{L}_{\rm full}=\mathcal{L}_\phi+\mathcal{L}_{\rm int}+\mathcal{L}_{\mathcal{X}}$ we mean the embedding of $\mathcal{L}_\phi$ into a consistent theory of particle physics. Here $\mathcal{L}_{\mathcal{X}}$ summarises all terms in the Lagrangian that do not contain $\phi$. It comprises the SM Lagrangian. Obviously $\mathcal{L}_{\rm infl} \subset \mathcal{L}_\phi \subset \mathcal{L}_{\rm full}$.
\end{itemize}
An important quantity for cosmology is the temperature at the onset of the radiation dominated era, which is usually referred to as \emph{reheating temperature} $T_R$. 
In principle it is not clear whether such a temperature can be defined because reheating tends to produce particles with highly non-thermal momentum distributions, and it is not clear that rescatterings can drive these to an equilibrium distribution (that allows to define a temperature) by the time that $\rho_\gamma=\rho_\varphi$.
Therefore $T_R$ should rather be thought of as an effective parameter that characterises the energy density. If the radiation is in kinetic equilibrium, then the notion of an effective temperature $T$ can even provide an approximate measure for the occupation number of different modes. Since kinetic equilibration usually happens faster than thermal equilibration \cite{Ellis:1987rw,Dodelson:1987ny,Enqvist:1990dp,Enqvist:1993fm,McDonald:1999hd,Davidson:2000er,Garbrecht:2008cb,Berges:2010zv,Mazumdar:2013gya,Harigaya:2013vwa,Harigaya:2014waa,Mukaida:2015ria}, this may be justified for a large class of scenarios.
The reheating era approximately ends when the rate $\Gamma_\varphi$ at which $\varphi$ looses energy due to dissipation equals the Hubble rate $H$. 
$\Gamma_\varphi$ and $H$ both depend on $T$, and $T_R$ can be determined by solving the equality $\Gamma_\varphi=H$ for $T$.
A very simple expression can be found if $\Gamma_\varphi$ is independent of $T$, i.e. $\Gamma_\varphi=\Gamma_0={\rm const}$. 
Using the expression $H=\sqrt{8\pi^3 g_*/90}T^2/M_P$ for the Hubble rate during radiation domination, one finds
\begin{equation}
T_R=\sqrt{\Gamma_0 M_P }\left(\frac{90}{8\pi^3g_*}\right)^{1/4}\label{TR}
\end{equation}
Note that $T_R$ in any case is usually not the highest temperature during reheating \cite{Giudice:2000ex,Drewes:2014pfa}.

\subsection{Equation of motion for $\varphi$}
It is often assumed that $\varphi$ follows an equation of motion of the form
\begin{equation}\label{EOM}
\ddot{\varphi} + (3H + \Gamma_\varphi)\dot{\varphi} + \partial_\varphi \V(\varphi)=0.
\end{equation}
Here $H=\dot{a}/a$ is the Hubble rate, $a$ the scale factor and  
$\V(\varphi)$ the effective potential for $\varphi$, which includes all quantum and thermodynamic corrections. $\Gamma_\varphi$ is an effective dissipation rate that leads to the transfer of energy from $\varphi$ to $\mathcal{X}$. 
The statement that the dynamics can be described by a Markovian effective equation of motion of the type (\ref{EOM}) is not trivial. 
Both, inflation and reheating, take place in a highly non-trivial background. During inflation, fields reside in de Sitter spacetime. Reheating can occur almost instantaneously via an instability, in which case there may be no separation between macroscopic and microscopic time scales, as required for (\ref{EOM}) to be valid. 
Even in single field models where it is driven by coherent oscillations,
reheating is a complicated nonequilibrium process that involves a "background" of rapid $\varphi$-oscillations, the expansion of the universe and far from equilibrium states of the produced $\mathcal{X}$. In such an environment, the conventional S-matrix approach to quantum field theory cannot be applied. Starting point for a consistent derivation of (\ref{EOM}) is the nonequilibrium effective action \cite{Jackiw:1974cv,Calzetta:1986cq} in the Schwinger-Keldysh formalism \cite{Schwinger:1960qe,Bakshi:1962dv,Bakshi:1963bn,Keldysh:1964ud}.
There have been some re-derivations of equations of the type (\ref{EOM}) in the literature \cite{Morikawa:1986rp,Greiner:1998vd,Yokoyama:2004pf,Boyanovsky:2004dj,Anisimov:2008dz,Boyanovsky:2015xoa,Mukaida:2013xxa}, most recently \cite{Cheung:2015iqa}. In the following we will assume that (\ref{EOM}) holds during reheating and discuss the individual terms.

The effective potential $\V(\varphi)$ that $\varphi$ is exposed to differs from  $V(\phi)$. It includes quantum and thermodynamic corrections. Via these effects, 
$\V(\varphi)$ in principle is sensitive to all parameters in $\mathcal{L}_{\rm full}$ (primarily of course to those in $\mathcal{L}_{\phi}$).
Corrections threaten do spoil the flatness of a classically shallow potential that is required for a sufficiently long "slow roll" phase. 
On one hand, this could be corrections due to couplings to matter.
One can assume $\lambda\ll\alpha$ and $\varphi\ll m_\phi/\lambda$ to avoid this.
On the other hand, there are corrections from gravity.
Many models predict trans-Planckian field values $\varphi\gg M_P$ during inflation, where $M_P$ is the Planck mass.\footnote{This does not necessarily mean that the laws of physics as we know it (more precisely: the framework of quantum field theory) entirely brake down as long as physical observables take values below $M_P$,
but it does mean that the effective field theory approach of inflation, which is based on an expansion in powers of $\varphi/M_P$, breaks down. That is, even if quantum field theory remains a good description of nature in this regime (which is a brave extrapolation from the regime where it has been tested), there may be no practically calculable way to relate the full effective potential $\V(\varphi)$  to $V(\phi)$ and other parameters $\lambda, \alpha$ in $\mathcal{L}_{\rm full}$.}
During reheating, $\varphi$ usually takes values $\varphi<M_P$, and there is no trans-Planckian issue. However, in the presence of the primordial plasma formed by the inflaton's decay products,
 $\V(\varphi)$ may receive considerable thermal corrections, see \cite{Cheung:2015iqa} for a recent discussion.

The dissipative term $\Gamma_\varphi$ \emph{necessarily} appears in an interacting quantum field theory. This should be obvious from the basic laws of statistical mechanics: If the energy in an interacting system is concentrated in one degree of freedom, the interactions will redistribute it as the system approaches equilibrium.\footnote{Note that there is no fluctuation-term on the RHS of (\ref{EOM}), as one might expect from a generalised fluctuation-dissipation theorem. This is because the definition of $\varphi$ includes an average over thermal fluctuations. Near the potential minimum these average to zero. The equation of motion for $\phi$ would contain a noise term, see e.g. \cite{Greiner:1998vd,Yokoyama:2004pf,Boyanovsky:2004dj,Anisimov:2008dz,Cheung:2015iqa}.}
Usually $\Gamma_\varphi$ is thought to be negligible during inflation.\footnote{See \cite{Berera:2008ar} for a different viewpoint.}
The dissipation is, however, the driving force behind reheating. 
Inflation roughly ends when the $\partial_\varphi\V(\varphi)$ term in (\ref{EOM}) exceeds $3H\dot{\varphi}$. %(while $\Gamma_\varphi\dot{\varphi}$ remains negligible). 
This marks the beginning of the reheating era, during which $\varphi$ oscillates around its minimum. 
At this time one typically observes the hierarchy $\Gamma_\varphi\dot{\varphi}\ll 3H\dot{\varphi} \lesssim \partial_\varphi\V(\varphi)$.
The relative smallness of $\Gamma_\varphi\dot{\varphi}$ implies that $\varphi$ looses only a small fraction of its energy per oscillation. However, the total amount of energy that is transferred from $\varphi$ to $\mathcal{X}$ is usually the largest at early times because $\rho_\varphi$ is huge in the beginning (and redshiftet a later times). 

The end of the reheating era comes when $\Gamma_\varphi\simeq H$, which leads to a rapid transfer of energy from $\varphi$ to $\mathcal{X}$ within a few Hubble times. 
Practically it is, however, highly nontrivial to determine this moment because $\V(\varphi)$ and $\Gamma_{\varphi}$ are time dependent. 
One one hand the time dependence comes from the fact that interactions with the oscillating background field $\varphi$ lead to 
$\varphi$-dependent (and thus time-dependent) effective particle masses. On the other hand there is a back-reaction of the produced particles on the dynamics of $\varphi$: for bosonic $\mathcal{X}_i$, induced transitions increase the dissipation rate if the modes in the final state are already occupied, for fermionic $\mathcal{X}_i$ in the final state there is a suppression due to Pauli blocking. 
The combination of these effects can lead to the well-known phenomenon of \emph{parametric resonance} \cite{Traschen:1990sw,Kofman:1994rk,Kofman:1997yn}.
In the most efficient version of the resonance, particles are directly produced from the non-adiabatic background formed by $\varphi$ (rather than in processes involving individual $\phi$-quanta), and the rate of dissipation has to be determined by nonperturbative methods.

It is obvious that $\Gamma_\varphi$ must depend on the parameters in $\mathcal{L}_\phi$, in particular $\lambda$, regardless of whether it is calculated by perturbative or nonperturbative techniques. 
What is important in the present context is that $\Gamma_\varphi$ in general also depends on the parameters in $\mathcal{L}_{\mathcal{X}}$, i.e. on $\alpha$ and the particle masses in the plasma.  
In the present work, we use a number of representative examples to illustrate how this dependence arises. 
In section \ref{WhatIsRSec} we define the parameter $\Rrad$, which parametrises the effect of the expansion history during reheating on the CMB.
$\Rrad$ depends on $\V(\varphi)$ and $\Gamma_\varphi$.
In section \ref{DepOnModParamSec} we use four simple examples to illustrate the dependence of $\Gamma_\varphi$ and $\Rrad$ on the parameters in $\mathcal{L}_\phi$ and $\mathcal{L}_{\mathcal{X}}$. 
In section \ref{SecDiscussion} we discuss how this dependence affects the questions 1)-4).
In section \ref{Conclusions} we conclude.

\section{How the reheating era affects the CMB}\label{WhatIsRSec}
The details of the post-inflationary era do not directly affect the spectrum of temperature fluctuations in the CMB at the time of photon decoupling, see e.g. \cite{Martin:1997zd}.\footnote{The observed CMB is of course affected by a wide range of phenomena that occurred after photon decoupling, as the light had to pass through  gravitational potentials and matter on the way to us.}
The expansion history does, however, affect the way how physical scales at present time and during inflation relate to each other. The reheating era is the least understood part of the post-inflationary history.

\subsection{The parameter $\Rrad$}  
In this work, we entirely focus on the effect that reheating has on the CMB due to the expansion history during the reheating era. This effect can be quantified by a single parameter $\Rrad$ and is very generic, i.e.,\ unavoidable.\footnote{Other possible signatures of reheating in the CMB that have been studied include non-Gaussianities \cite{Bond:2009xx,Cicoli:2012cy} and curvature-perturbations \cite{Ringeval:2013hfa,Mazumdar:2014haa,Mazumdar:2015xka}, but these are more model dependent.}
If one assumes that no significant amount of entropy is injected into the bath of radiation after the end of the reheating era,\footnote{If additional entropy is injected into the plasma after reheating, one can still define a parameter $\Rrad$, but the contributions from different reheating eras cannot be disentangled.}
the set of equalities \cite{Martin:2006rs}
\begin{eqnarray}
1+ \zend  &=& \frac{\areh}{\aend}(1+\zreh) =
\frac{\areh}{\aend} \left(\frac{\rhoreh}{\rhotildegamma}
\right)^{1/4} 
%\\  &
=
%& 
\frac{\areh}{\aend}
\frac{\rhoreh^{1/4}}{\rhoend^{1/4}}
\frac{\rhoend^{1/4}}{\rhotildegamma^{1/4}} \equiv
\frac{1}{\Rrad} \left(\frac{\rhoend}{\rhotildegamma}
\right)^{1/4}
\label{preRradDef}
\end{eqnarray}
holds.
Here $\rhoend$ is the energy density at the end of inflation and $\rhotildegamma$ the radiation density at late times.
That is, the effect of the reheating era on the redshifting of modes can be parametrised by the integrated observable\begin{equation} 
\label{RradDef}
R_{\rm rad}=\left(\frac{\rhoend}{\rhoreh}\right)^{1/4}\frac{\aend}{\areh}.
\end{equation}
The variable $R_{\rm rad}$ essentially characterises the deviation of the expansion history from a radiation dominated universe.
If the equation of state during reheating were $w=1/3$, then $R_{\rm rad}=1$.
More generally, $R_{\rm rad}$ can be calculated as \cite{Martin:2010kz}
\begin{eqnarray}\label{someRformula}
\ln R_{\rm rad} = \frac{\Delta N}{4}\left(3\wrehbar-1\right).
\end{eqnarray}
Here $\wrehbar$ is the average equation of state during reheating and $\Delta N$ the number of e-folds from the end of inflation until the end of reheating.
This relation was derived by inserting 
\begin{eqnarray}\label{waveraging}
\frac{\rhoend}{\rhoreh}=\exp\left[
\int_{N_{\rm end}}^{N_{\rm reh}} 
dN (1+\wreh)
\right]
=\exp\left[
(1+\wrehbar)\Delta N
\right]
\end{eqnarray}
into the definition (\ref{RradDef}).
%shows that N*
By using (\ref{waveraging}) again, one can remove $\Delta N$ from (\ref{someRformula}) and obtains
\begin{eqnarray}
\label{Rrad}
\ln \Rrad=
\frac{1-3\wrehbar}{12\left(1+\wrehbar\right)}
\ln\left(\frac{\rhoreh}{\rhoend}\right)\,.
\end{eqnarray}
The relations between $\zend$, $\rhoend$ and $\Rrad$ given by (\ref{preRradDef}) and (\ref{Rrad}) can be understood qualitatively by considering the example that $\varphi$ oscillates in a harmonic potential $\V(\varphi)\propto \varphi^2$ during reheating. In this case the averaged equation of state is $\wrehbar=0$, just like in a matter dominated era, leading to a redshift $\rho \propto a^{-3}$.
This of course implies $a\propto \rho^{-1/3}$, while one would have $a\propto \rho^{-1/4}$ during radiation domination. Therefore the amount of redshifting (increase in $a$) that a given CMB more experiences while $\rho$ decreases by a fixed amount is larger during the reheating era than during radiation domination. 
For fixed $\rhoend$ one expects a larger $\zend$ if there was a reheating era with $\wrehbar=0$, compared to a scenario in which the universe immediately enters a radiation dominated era with $\wrehbar=1/3$ after inflation.\footnote{Note that this is specific to the case of oscillations in a potential $\V(\varphi)\propto \varphi^p$ with $p<4$, cf. (\ref{PowerLawPotential}) and (\ref{powerlaww}).
For $p>4$ there is actually \emph{less} redshifting during reheating, leading to $\Rrad>1$ and a reduced $\zend$, see (\ref{RRange}).
For $p=4$ the equation of state is $\wrehbar=1/3$. As far as the expansion history is concerned, an era in which $\rho$ is dominated by $\varphi$-oscillations in a quartic potential is therefore indistinguishable from a radiation dominated era, and (\ref{Rrad}) gives $\Rrad=1$ irrespectively of $\rhoreh/\rhoend$. 
However, this does not mean that reheating did not leave any imprint in the CMB because usually there is a period of ``fast roll'' between the moment when inflation ends ($w=-1/3$) and the beginning of the oscillations ($\wrehbar=1/3$), which leads to a deviation from $\Rrad=1$.
} 
From (\ref{Rrad}) it is clear that the latter scenario leads to $\Rrad=1$ (because $\rhoreh=\rhoend$), while the reheating era (i.e.\ $\rhoreh<\rhoend$) leads to $\Rrad<1$,  hence (\ref{preRradDef}) indeed gives a larger $\zend$ for fixed $\rhoend$. 
This effect should be larger if the duration of the reheating era is longer. The duration of the reheating period is governed by the parameters $\lambda$: larger inflaton couplings imply a more efficient energy transfer from $\varphi$ to other degrees of freedom, leading to a shorter reheating period. Hence, one would expect that smaller $\lambda$ lead to a larger deviation of $\Rrad$ from unity and to a larger $\zend$ for given $\rhoend$. For the explicit example studied in section \ref{SimplePerturbaiveSec}, we indeed find $\zend\propto 1/\Rrad \propto \lambda^{-1/3}$, see (\ref{RexplicitResult}).

The precise relation between observed power spectrum, $\Rrad$ and quantum
fluctuation in the metric during inflation
is given in \cite{Martin:2010kz}, for the purpose of the present work it is sufficient to know that $R_{\rm rad}$ parametrises the effect of the expansion history during reheating on observable CMB-modes.
The great advantage of this method is that it appears to be rather model-independent.
The RHS of (\ref{Rrad}) depends on $\rhoend$, $\wrehbar$ and $\rhoreh$.
$\rhoend$ can usually be determined from the shape of $\V(\varphi)$ alone.
If reheating happens via coherent oscillations, then the equation of state $\wreh$ during reheating is (by definition) dominated by $\varphi$, 
which is determined by $\V(\varphi)$.
This seems realistic in large (single) field models with small $\lambda$; it is in general not true if reheating is driven by an instability, which can lead to turbulence and significant gradient energies during reheating.
In such models, $\wreh$ is in good approximation determined by $\V(\varphi)$ alone, and is independent of details of underlying particle physics model.
This suggests that one may be able to probe $\V(\varphi)$ in a ``model-independent'' way, i.e. without specification of its embedding into a model of particle physics.
However, the duration of reheating and the moment when it ends of course depends on the interactions of $\varphi$ with other degrees of freedom. This dependence enters $\rhoreh$. 
Therefore the RHS of (\ref{Rrad}) does not only depend on $\V(\varphi)$ alone, but at least on some of the inflaton's couplings to radiation and matter in $\mathcal{L}_\phi$ (the strength of which is characterised by $\lambda$). 
In general, $\Rrad$ also depends on the interactions of the plasma constituents $\mathcal{X}$ with each other (the strength of which is characterised by $\alpha$) and other parameters in $\mathcal{L}_{\mathcal{X}}$, especially during preheating. 
One in principle has to specify the full particle physics model $\mathcal{L}_{\rm full}$ that $\phi$ is embedded into to calculate $\Rrad$, which means that this parameter is only of limited use if one wants to constrain $\V(\varphi)$ during reheating from the CMB. This introduces a systematic uncertainty \cite{Kinney:2005in,Peiris:2006sj,Adshead:2010mc} in any extraction of $\V(\varphi)$-parameters from CMB observations. 
However, as we will see, the dependence of $\Rrad$ on the parameters in $\mathcal{L}_{\mathcal{X}}$ is usually very weak in models without a parametric resonance. In such scenarios one can use $\Rrad$ to constrain $\lambda$ and $T_R$.

\subsection{How to make use of $\Rrad$}
There are different ways to make use of the information contained in $\Rrad$. 
\paragraph{A) Refining particle physics model predictions} - The most obvious approach to use $\Rrad$ is to refine predictions for the CMB power spectrum in a given particle physics theory $\mathcal{L}_{\rm full}$.
It is often assumed that the spectrum of temperature fluctuations at the time of photon decoupling is determined by $\V(\varphi)$ alone.
From the previous considerations it is clear that the redshifting during the reheating phase, which can last many $e$-folds, affects the overall scaling of modes. 
Hence, specification of $\V(\varphi)$ alone is not enough to predict the pattern of CMB perturbations, but also interactions of $\phi$ and the primordial plasma play a role.
However, one specifies a complete theory $\mathcal{L}_{\rm full}$, then $\V(\varphi)$ and $\Rrad$ are at least in principle calculable (though this can be hard in practice).
This allows to eliminate the systematic uncertainty coming from $\Rrad$.

\paragraph{B) Constraining $T_R$ in a given inflationary model} - Often one does not want to specify the entire particle physics theory $\mathcal{L}_{\rm full}$.
Instead, one could specify only the inflationary model $\V(\varphi)$ and pose the question whether a particle physics model that accommodates this $\V(\varphi)$ allows to reproduce the observed CMB.
For this purpose it is not helpful to follow the approach in the previous paragraph and try all (infinitely many?) particle physics models that may accommodate $\V(\varphi)$ one by one.
Instead, one can ask if there exist any $\mathcal{L}_{\mathcal{X}}$-independent constraints on $\Rrad$.
These indeed exist and have been used in \cite{Martin:2010kz,Cook:2015vqa}.
The positive energy theorem imposes 
\begin{equation}\label{boundaryw1}
\wrehbar<1.
\end{equation} 
Furthermore 
\begin{equation}
\wrehbar>-1/3, 
\end{equation}
as otherwise one would be in a period of inflation, which by definition is different from reheating.
Moreover, obviously reheating happened after inflation,
\begin{equation}\label{boundaryend}
\rhoreh<\rhoend,
\end{equation}
and before big bang nucleosynthesis, 
\begin{equation}\label{boundarynuc}
\rho_{\rm BBN}<\rhoreh.
\end{equation}
Hence, $\Rrad$ cannot take arbitrary values. 
The above constraints are model independent.
Let us assume that the predictions of the model $\V(\varphi)$ can only be made consistent with CMB observations if the unknown parameter $\Rrad$ lies in an interval $[\Rrad^{\rm min},\Rrad^{\rm max}]$ of observationally allowed values.
Then this implies 
\begin{eqnarray}\label{Trconstraint}
(\Rrad^{\rm min})^{\frac{12(1+\wrehbar)}{1-3\wrehbar}} < \frac{\rhoreh}{\rhoend} < (\Rrad^{\rm max})^{\frac{12(1+\wrehbar)}{1-3\wrehbar}}
\end{eqnarray}
and imposes a constraint on $\rhoreh$.
Once $\V(\varphi)$ is specified near its minimum, one can impose even stronger constraints on $\Rrad$ because $\wrehbar$ and $\rhoend$ can be calculated, and $\Rrad$ in (\ref{Rrad}) is a function of $\rhoreh$ alone.
For example, for a power law potential
\begin{equation}\label{PowerLawPotential}
\V(\varphi)\propto (\varphi/M_P)^p
\end{equation}
one finds that during the $\varphi$-oscillations \cite{Turner:1983he}\footnote{Note that (\ref{powerlaww}) might not be applicable during the whole reheating era because there can be a phase of ``fast roll'' between the end of inflation and the beginning of the oscillations. We ignore this technical detail for simplicity; it does not change the discussion qualitatively because $w$ during such a period can also be calculated from $\V(\varphi)$ alone. }
\begin{equation}\label{powerlaww}
\wrehbar=\frac{p-2}{p+2}
\end{equation}
if one assumes that $\rho=\rho_\varphi$. Deviations from this due to the presence of a small radiation component are small \cite{Martin:2010kz}, hence
\begin{equation}
\Rrad=\left(\frac{\rhoreh}{\rhoend}\right)^{\frac{4-p}{12p}}
\end{equation}
holds in good approximation.
This yields 
\begin{eqnarray}\label{RRange}
 \left(\frac{\rho_{\rm BBN}}{\rhoend}\right)^{\frac{4-p}{12p}} &<& \Rrad < 1 \ {\rm for } \  p\leq 4,\\
\left(\frac{\rho_{\rm BBN}}{\rhoend}\right)^{\frac{4-p}{12p}} &>& \Rrad > 1 \ {\rm for } \  p> 4
\end{eqnarray}
and imposes a bound on $\rhoreh$
\begin{eqnarray}
(\Rrad^{\rm min})^{\frac{12p}{4-p}} < \frac{\rhoreh}{\rhoend} &<& (\Rrad^{\rm max})^{\frac{12p}{4-p}} \ {\rm for } \  p\leq 4,\\
(\Rrad^{\rm min})^{\frac{12p}{4-p}} > \frac{\rhoreh}{\rhoend} &>& (\Rrad^{\rm max})^{\frac{12p}{4-p}} \ {\rm for } \  p> 4.\label{Trconstraintp}
\end{eqnarray}
This window can be compared to the window allowed by the constraints (\ref{boundaryend}) and (\ref{boundarynuc}).
For $\V(\varphi)\simeq M_\varphi^2\varphi^2/2$ and $\varphi\simeq M_P$ at the end of inflation, we can use (\ref{boundaryend}) and (\ref{boundarynuc}) with $\rho_{\rm BBN}\simeq(10 {\rm MeV})^4$ to find 
\begin{eqnarray}\label{explicitBBNconstraint}
10^{-4} \left(\frac{\rm GeV}{M_\varphi}\right)^{1/6} < \Rrad < 1,
\end{eqnarray}  
which can be compared to the observationally allowed window $\Rrad^{\rm min}  <\Rrad < \Rrad^{\rm max}$.
If there is an overlap, this allows to put a bound on $\rhoreh$ and the reheating temperature.
If there is no overlap, then it even allows to exclude the inflationary model $\V(\varphi)$.
If one makes the assumption that the equation of state $w$ is approximately constant during the reheating era, these considerations can be translated into predictions for the scalar power spectrum amplitude and spectral index \cite{Cook:2015vqa}.

\paragraph{C) Bayesian approach} - Finally, one may consider $\Rrad$ a free parameter in Bayesian fits, in addition to those parameters appearing in  $\V(\varphi)$ 
and the usual free parameters of the $\Lambda$CDM model (as well as other parameters that characterise the late time physics). 
For practical purposes, it turns out to be better to use ${\rm R}_{\rm reh}\equiv\Rrad\rhoend^{1/4}/M_P$. The relations (\ref{boundaryw1}) - (\ref{boundarynuc}) can be used to choose a well-motivated prior (for instance a flat prior in $\ln {\rm R}_{\rm reh}$ within the allowed region). 
If the posterior is more peaked than prior, then on has obtained non-trivial information about the reheating stage \cite{Martin:2014nya}. 
Similar ideas to include the post-inflationary history into Bayesian fits have previously been discussed in \cite{Mortonson:2010er,Easther:2011yq}.

\section{Dependence of $\Rrad$ on model parameters}\label{DepOnModParamSec}
Let us consider the expression (\ref{Rrad}) for $\Rrad$. 
In the scenarios we consider here, the averaged equation of state $\wrehbar$ mainly depends on $\V(\varphi)$ because 
the total energy density $\rho=\rho_\varphi+\rho_\gamma$ during reheating is dominated by $\rho_\varphi$, and the equation of state for $\varphi$ during oscillations near the minimum of $\V(\varphi)$ is determined by the shape of the effective potential in that region, cf. (\ref{PowerLawPotential})-(\ref{powerlaww}).  
Hence, $\wrehbar$ is in good approximation a property of the inflationary model $\V(\varphi)$ alone and independent of its embedding into a particle physics framework.\footnote{Here we continue to ignore the fact that the relation between the tree level potential $V(\phi)$ in $\mathcal{L}_{\rm infl}$ and the effective potential $\V(\varphi)$ in principle via loop corrections depends on all parameters in $\mathcal{L}_{\rm full}$, and that therefore $\V(\varphi)$ and $\Rrad$ strictly speaking cannot be treated as independent entities.}
We  continue to assume that the comoving entropy is in good approximation conserved after reheating. 
This assumption 
in principle brings in a dependence on the choice of particle physics models $\mathcal{L}_{\rm full}$ because it e.g. excludes models that use the decay of a heavy particle to dilute unwanted relics 
and theories containing moduli that decay at late time \cite{Kane:2015jia}.

Under these assumptions,  the only dependence of $R_{\rm rad}$ on parameters other than those in $\V(\varphi)$ comes from $\rhoreh$, i.e. the energy density at the moment when the reheating era ends and the universe becomes radiation dominated.
This moment of course depends on the mechanism by which $\varphi$ dissipates its energy, which depends on the interactions between $\varphi$ and other degrees of freedom. 
It can roughly be estimated as the moment when $\Gamma_\varphi=H$.
Therefore, the dependence of $\Rrad$ on the details of the particle physics framework is in good approximation determined by the effective damping parameter $\Gamma_\varphi$.
Many studies of reheating in a particle physics context are based on the assumptions 
\begin{itemize}
\item[$i)$] $\Gamma_\varphi$ is constant during the reheating phase.
\item[$ii)$] $\Gamma_\varphi$ only depends on the parameters in $\mathcal{L}_\phi$. 
\end{itemize}
Both of these assumptions are crucial when translating bounds on $\Rrad$ into constraints on particle physics models, but both of them are in general not justified.
In the following, we use a number of simple examples 
that have been studied in the past literature to illustrate different effects that can violate $i)$ and $ii)$.

\subsection{Perturbative reheating without feedback}\label{SimplePerturbaiveSec}
In the simplest case, $\varphi$ slowly dissipates its energy perturbatively by decays of individual inflaton quanta into much lighter particles. 
This happens if $\lambda$ are sufficiently small that the effective mass of $\mathcal{X}$-modes (which oscillates due to its coupling to $\varphi$) changes only adiabatically, and if the occupation numbers in the primordial plasma at all times remain so small that the feedback of produced $\mathcal{X}$-particles on the $\varphi$ evolution is negligible.
In this case assumptions $i)$ and $ii)$ are both fulfilled,
and one may approximate $\Gamma_\varphi$ by the vacuum decay rate $\Gamma_0$ of $\phi$ particles, which is constant throughout the reheating era.

\paragraph{Equations of motion} - 
Under the above assumptions the decay is much slower than the $\varphi$-oscillations, and instead of (\ref{EOM}) one may solve a coupled set of equations of motion for $\rho_\varphi$ and $\rho_\gamma$ averaged over a few oscillations. 
If the effective potential is, for instance, quadratic near the minimum,
\begin{equation}
\V(\varphi)\simeq \frac{M_\phi^2}{2}\varphi^2,
\end{equation} 
then (\ref{powerlaww}) implies that $\rho_\varphi$ redshifts like matter. Moreover, for sufficiently small $\alpha$, the produced radiation can thermalise faster than the duration of the reheating process \cite{Mukaida:2015ria}, and may assume that the $\mathcal{X}$-occupation numbers can be characterised by an effective temperature $T$ in some approximation.
Then one can use the set of equations 
\begin{eqnarray}
\frac{d \rho_\phi}{d t}+3H\rho_\phi+\Gamma_\varphi\rho_\varphi&=&0\label{BE1}\\
\frac{d \rho_\gamma}{d t}+4H\rho_\gamma-\Gamma_\varphi\rho_\varphi&=&0\label{BE2}
\end{eqnarray}
with
\begin{equation}\label{RHOofT}
\rho_\gamma=\frac{\pi^2 g_*}{30}T^4.
\end{equation}
It is convenient to introduce the variables $\Phi\equiv\rho_\varphi a^3/m_\phi$, $R\equiv\rho_\gamma a^4$ and $x\equiv a m_\phi$, in terms of which we can rewrite (\ref{BE1}) and (\ref{BE2}) as
\begin{eqnarray}
\frac{d\Phi}{dx}&=&-\frac{\Gamma_\varphi}{Hx}\Phi\label{be1}\\
\frac{dR}{dx}&=&\frac{\Gamma_\varphi}{H}\Phi\label{be2}
\end{eqnarray}
with
\begin{eqnarray}
H&=&\left(\frac{8\pi}{3}\right)^{1/2}\frac{m_\phi^2}{M_P}\left(\frac{R}{x^4}+\frac{\Phi}{x^3}\right)^{1/2}\label{Hubble}\\
T&=&\frac{m_\phi}{x}\left(\frac{30}{\pi^2g_*}R\right)^{1/4}\label{TofR}.
\end{eqnarray}
With our conventions, $x$ essentially measures the scale factor $a$ in units of its value at the end of inflation, $x=a/a_{\rm end}$ 

\paragraph{Maximal temperature and reheating temperature} - We define the \emph{maximal temperature} $T_{MAX}$ as the maximum of $T$  as a function of $x$ and the \emph{reheating temperature} $T_R$ as the temperature in the moment $\rho_\phi=\rho_R$. The latter corresponds to 
\begin{equation}
\frac{R}{x^4}=\frac{\Phi}{x^3}.\label{PhiRadEq}
\end{equation} 
and coincides in good approximation  with the moment when $\Gamma_\varphi=H$. 
Prior to the moment $\Gamma_\varphi=H$ the fractional energy loss of $\varphi$ is negligible, and one can in good approximation treat $\Phi=\Phi_I$ as a constant and treat it as an external source in (\ref{BE2}). 
With the approximation $\Gamma_\varphi=\Gamma_0$ it is easy to obtain the solution 
\begin{equation}
R=A_0\frac{2}{5}(x^{5/2}-1)\label{R0}
,\end{equation}
where $A_0$ is given by the function
\begin{equation}
A_n\equiv\frac{\Gamma_n}{m_\phi}\sqrt{\Phi_I}\frac{M_P}{m_\phi}\left(\frac{30}{\pi^2g_*}\right)^{n/4}\left(\frac{3}{8\pi}\label{An}\right)^{1/2}
,\end{equation}
which we defined for arbitrary $n$ for later use.
From this expression and (\ref{TofR}) we can find $T$ as a function of $x$
\begin{equation}
T=m_\phi \left(A_0\frac{30}{\pi^2g_*}\frac{2}{5}\right)^{1/4}\left(x^{-3/2}-x^{-4}\right)^{1/4}\label{T1}
.\end{equation}
The maximum of (\ref{T1}) is at $x_{max}=(8/3)^{2/5}\simeq 1.48$, hence we find $T_{MAX}=T_{max}$, where
\begin{eqnarray}
T_{max}&=& m_\phi \left(A_0\frac{30}{\pi^2g_*}\frac{2}{5}\right)^{1/4}\left(x_{max}^{-3/2}-x_{max}^{-4}\right)^{1/4}\\
&\simeq&0.6 \left( \frac{\Gamma_0}{g_*} M_P \right)^{1/4}\V_I^{1/8} \simeq 0.7 T_R^{1/2}\left(\frac{\V_I}{g_*}\right)^{1/8}\label{Tmax}
.\end{eqnarray}
Here we have introduced $\V_I$ as the value of $\V(\varphi)$ at the beginning of reheating.
In simple large field models, this can be estimated as $\varphi\simeq M_P$, i.e.\ $\V_I=M_P^2m_\phi^2/2$.
The value of $T_R$ under the present assumptions is in good approximation given by (\ref{TR}) because most of the $\varphi$-energy is dumped into radiation almost instantaneously in the moment when $\Gamma_0=H$.
These results have first been found in \cite{Giudice:2000ex}  and have later been confirmed analytically and numerically in \cite{Drewes:2014pfa}.

\paragraph{A simple model} - 
As a concrete example, one may assume that $\phi$ couples only to another scalar field $\chi$ via interactions 
\begin{equation}\label{toymodel}
\mathcal{L}_{\rm int}=
\frac{\lambda_4}{3!}\phi\chi^3+\frac{\lambda_3}{2} m_\phi \phi\chi^2.
\end{equation}
Then the frequency of a $\chi$-mode $k$ is
\begin{equation}\label{OmegachiDef}
\Omega_\chi^2 = \omega_\chi^2 +\lambda_3 m_\phi \varphi  +\mathcal{O}[\lambda_3^2,\lambda_4^2],
\end{equation}
where $\omega_\chi^2\equiv\textbf{k}^2 + m_\chi^2$.  The terms of higher order in $\lambda_3$ and $\lambda_4$ include temperature- and momentum-dependent self-energy terms from the interaction with particles in the plasma. 
To ensure that a perturbative treatment is justified, the adiabaticity condition 
\begin{equation}\label{ADIABATICITY}
\frac{\dot{\Omega}_\chi}{\Omega_\chi^2}\ll 1
\end{equation}
has to be fulfilled. If we assume for simplicity that the curvature of $\V(\varphi)$ near its minimum is given by the vacuum mass $m_\phi$, i.e. $M_\varphi\simeq m_\phi$, then 
the frequency of $\varphi$-oscillations is roughly $m_\phi$, and the adiabaticity condition translates into the simple requirement
\begin{eqnarray}\label{adiabaticitysimple}
\lambda_3\varphi|_{x=1}\ll \frac{\omega_\chi^3}{m_\phi^2}.
\end{eqnarray}
This condition holds for all modes if $\lambda_3\varphi|_{x=1}<m_\phi$.
If we assume that the inflaton mass is much bigger than  $m_\chi$, then the vacuum damping rate $\Gamma_0$ is driven by are decays $\phi\rightarrow\chi\chi$ and $\phi\rightarrow\chi\chi\chi$. For $\lambda_3\simeq\lambda_4$, the $1\rightarrow 2$ decays dominate due to the larger phase space, and $\Gamma_0\simeq \lambda_3^2m_\phi/(32\pi)$ %SYMMETRY FACTOR!
is constant and given by a single coupling constant $\lambda_3$.
In this case $\Rrad$ only depends on the parameters in $\V(\varphi)$ and $\lambda_3$ and is calculable analytically. 
For example, for the simple %(but disfavoured \cite{Planck:2013jfk}) 
model $\V(\varphi)=\frac{1}{2}M_\varphi^2\varphi^2$, one obtains $\wrehbar=0$ from (\ref{powerlaww}). The ratio $\rhoreh/\rhoend$ in (\ref{Rrad}) can be determined from (\ref{TR}).
If one takes the usual estimate that reheating starts when $\varphi\simeq M_P$, then $\rhoend\simeq M_\varphi^2 M_P^2/2$.  
At $T=T_R$, (\ref{RHOofT}) and (\ref{TR}) yield $\rho_\gamma=\Gamma_0^2 M_P^2 \frac{3}{8\pi}\simeq \rhoreh/2$ and 
\begin{equation}
\frac{\rhoreh}{\rhoend}=\frac{3}{2\pi}\frac{\Gamma_0^2}{M_\varphi^2}.
\end{equation}
With $M_\varphi\simeq m_\phi$ the dependence on these parameters cancels and one obtains $\rhoreh/\rhoend\simeq 3\lambda_3^4(2\pi)^{-3}2^{-8}$ for this particular model.
Combining these results, we find
\begin{equation}\label{RexplicitResult}
\Rrad\simeq \lambda_3^{1/3}\frac{6^{1/12}}{2\pi^{1/4}}\simeq  0.4\lambda_3^{1/3}.
%\lambda_3^{1/3}\frac{3^{1/12}}{(2\pi)^{1/4}\sqrt{2}}\simeq  \frac{\lambda_3^{1/3}}{2}.
\end{equation}
This result of course only holds for $\lambda_3\ll 1$.
Therefore, in this particular model with the above assumptions, $i)$ and $ii)$ are fulfilled, and a constraint on $\Rrad$ directly constrains the inflaton coupling. 
For instance, (\ref{explicitBBNconstraint}) immediately implies $1\gg \lambda_3>10^{-11}/\sqrt{m_\phi/{\rm GeV}}$. Stronger bounds can be derived if constraints on $\Rrad$ from CMB observations are taken into account.
The relation between $\Rrad$ and $\lambda_3$ can be used to 
express $\Gamma_0$ in terms of $\Rrad$ and estimate the reheating temperature as 
\begin{equation}\label{TRestimate}
T_R\simeq 
\frac{\sqrt{m_\phi M_P}}{g_*^{1/4}}
\Rrad^3.
\end{equation}

\subsection{Induced transitions and thermal feedback}\label{NarrowResonanceSec}
The calculation in the previous section \ref{SimplePerturbaiveSec} assumed that $\Gamma_\varphi$ is simply given by the vacuum decay rate of $\phi$-particles.
In reality, the inflaton does not decay in vacuum, and there is a feedback of the produced particles on $\Gamma_\varphi$. 
This effect is crucial and can lead to a resonant production of particles.
In the simple toy model (\ref{toymodel}), for example, the $1\rightarrow 2$ decay produces significant amounts of $\chi$-particles with momenta $|\textbf{k}|=m_\phi/2$. 
This particular mode will reach a high occupation number $f_\chi(m_\phi/2)$. 
As soon as $f_\chi(m_\phi/2)$ is of order unity, one cannot neglect the effect of Bose enhancement on $\Gamma_\varphi$ any more. Then the approximation $i)$ of constant $\Gamma_\varphi$ certainly breaks down, and the effective decay rate is\footnote{The opposite effect would be achieved if the inflaton couples exclusively to fermions. In that case, one would find a Pauli suppression instead of (\ref{NarrowResonanceGamma}), possibly leading to a lower $\rhoreh$.}
\begin{equation}\label{NarrowResonanceGamma}
\Gamma_\varphi\simeq \Gamma_0[1+2 f_\chi(m_\phi/2)],
\end{equation}
How big the occupation number $f_\chi(m_\phi/2)$ gets is determined by the interplay of $\chi$-production, redshifting (which reduces $f_\chi(m_\phi/2)$ by changing the produced particles' momenta), inelastic scatterings and $\chi$-decays (both of which obviously reduce $f_\chi(m_\phi/2)$).
Let us consider two extreme cases.

\paragraph{Very small $\alpha$: negligible interactions} -  
If the $\chi$-interactions are very weak ($\alpha\ll\lambda$), then one may argue that inelastic scatterings and decays of $\chi$-particles are negligible. 
Since (\ref{NarrowResonanceGamma}) does not directly depend on $\alpha$, assumption $ii)$ still holds in this case. 
It is straightforward to solve the coupled set of equations for $\varphi$ and the distribution function $f_\chi(\k)$.
For $\varphi$ one can use (\ref{EOM}) with (\ref{NarrowResonanceGamma}), but for the $\chi$-particles one cannot use the momentum-integrated rate equation (\ref{BE2}) because one needs to distinguish $|\k|=m_\phi/2$ from other modes.
However, numerical evaluation is straightforward and allows to directly translate a constraint on $\Rrad$ into a constraint on $\lambda$ and $T_R$.
If $\lambda$ is large enough, the particle production enters a so-called \emph{narrow resonance} due to Bose enhancement.

\paragraph{Large $\alpha$: instant thermalisation } 
For $\alpha\neq0$, the occupation numbers of the produced particles are driven to  thermal equilibrium by their interactions. This is a highly non-trivial process \cite{Ellis:1987rw,Dodelson:1987ny,Enqvist:1990dp,Enqvist:1993fm,McDonald:1999hd,Davidson:2000er,Berges:2010zv,Mazumdar:2013gya,Harigaya:2014waa,Mukaida:2015ria}.
The time scale of equilibration varies vastly, it can be either longer or shorter than the duration of reheating \cite{Mukaida:2015ria}.
Let us for illustrative purposes assume that this process is very efficient ($\alpha\gg\lambda$), and the produced particles thermalise almost instantaneously (compared to the duration of the reheating era). 
Then the occupation numbers in the plasma at all times are characterised by a time dependent effective temperature $T$. In this case, there is no narrow resonance because the occupation number $f_\chi(m_\chi/2)$ is rapidly reduced to the equilibrium value $f_B(m_\phi/2)$, where $f_B$ is the Bose-Einstein distribution.
However, there is still Bose enhancement. 
The instantaneous thermalisation approximation may be unrealistic, but is "conservative" in the sense that it probably gives the smallest possible feedback of the radiation density on $\Gamma_\varphi$ for given $\alpha$. 
The rate $\Gamma_\varphi$ in this case can be calculated by means of thermal quantum field theory. A detailed studies in \cite{Drewes:2013iaa,Drewes:2015eoa} have shown that $\Gamma_\varphi$ in general has a complicated dependence on $T$, the shape of which strongly depends on the type of interactions in the primordial plasma and their strengths $\alpha$. It is convenient to parametrise $\Gamma_\varphi$ as
\begin{equation}\label{GammaExpansionInn}
\Gamma_\varphi=\sum_{n=0}^\infty \Gamma_n\left(\frac{T}{m_\phi}\right)^n 
\end{equation}
because $\Phi=\Phi_I$ is approximately constant prior to $\Gamma_\varphi=H$ and (\ref{be2}) can be solved analytically if $\Gamma_\varphi$ as a function of $T$ can locally be approximated by a monomial $\Gamma_\varphi\simeq \Gamma_n (T/m_\phi)^n$, 
\begin{equation}
R=\left(A_n\frac{1-n/4}{5/2-n}\left( x^{5/2-n} - x_i^{5/2-n} \right) + R_i^{1-n/4}\right)^{1/(1-n/4)}.\label{Rn}
\end{equation}
Here $R_i$ is $R$ at some initial moment $x=x_i$. This allows to construct a general solution piecewise as long as the polynomial  (\ref{GammaExpansionInn}) can locally be approximated by a monomial function.

For simplicity, we consider a situation where 
\begin{equation}\label{GammaSecondPower}
\Gamma_\varphi\simeq \Gamma_0 + \Gamma_2 \frac{T^2}{m_\phi^2}
\end{equation}
is a good approximation. This behaviour can e.g.\ be observed in the toy model (\ref{toymodel}) at low and intermediate temperatures if we set $\lambda_3=0$ to avoid a violation of the condition (\ref{adiabaticitysimple}).
Then $\Gamma_\varphi$ can be approximated by  \cite{Drewes:2013iaa}
\begin{eqnarray}\label{GammaApproxModel}
\Gamma_\varphi &\simeq& %\theta(T_2-T)
\frac{\lambda_4^2 m_\phi}{768\pi}\left(
\frac{1}{(2\pi)^2} + \frac{T^2}{m_\phi^2}
\right).
\end{eqnarray}
We have assumed that transport is entirely driven by exchange of screened one-particle states, and collective \emph{luon} excitations \cite{Drewes:2013bfa} play no role. 
The origin of the different terms in (\ref{GammaApproxModel}) is discussed in \cite{Drewes:2013iaa,Drewes:2014pfa}. 
the first term is given by decays $\phi\rightarrow\chi\chi\chi$ and the second term comes from scatterings $\phi\chi\rightarrow\chi\chi$ (and their inverse).
This behaviour is observed for $M_\chi\lesssim M_\phi\simeq m_\phi$ and is consistent with the results found in \cite{Parwani:1991gq}. Here $M_\chi$ is the effective thermal mass of $\chi$-particles in the plasma.
This approximation (\ref{GammaApproxModel}) breaks down when $M_\chi$ becomes larger than $M_\phi\simeq m_\phi$.
As an example, we consider
\begin{equation}\label{Lchi}
\mathcal{L}_{\mathcal{X}}=\frac{1}{2}\partial_\mu\chi\partial^\mu\chi-\frac{1}{2}m_\chi^2\chi^2 - \frac{\alpha}{4!}\chi^4.
\end{equation}
Apart from driving the $\chi$-particles to equilibrium, this interaction also dresses them with a thermal mass
\begin{equation}
M_\chi^2\simeq m_\chi^2 + \frac{\alpha}{24}T^2.\label{ThermalMass}
\end{equation}
from forward scatterings. In this case, the approximation (\ref{GammaApproxModel})  holds for $T<m_\phi/\sqrt{\alpha}$.

In this or any other model where (\ref{GammaSecondPower}) is a good approximation, an analytic solution for the equations (\ref{be1})-(\ref{TofR}) can be found. The behaviour depends on whether the temperature in the early universe exceeds a the critical value 
\begin{equation}
T_1\equiv m_\phi(\Gamma_0/\Gamma_2)^{1/2}
\end{equation}
  at which finite temperature effects start to dominate $\Gamma_\varphi$. 
If so, this happens at the moment when $x=x_1$ and $R=R_1$, with $1<x_1<x_{max}\simeq 1.48$.
Hence, we can expand (\ref{T1}) around $x=1$ to find 
\begin{eqnarray}
x_1&\simeq&1 +\left(\frac{\Gamma_0}{\Gamma_2}\right)^{1/2}\frac{\pi^2g_*}{30A_0} \ , \ R_1=A_0\frac{2}{5}(x_1^{5/2}-1) 
.\end{eqnarray}
One can distinguish three qualitatively different cases.
\begin{itemize}
\item If $T_R<T_{max}<T_1$, where $T_{max}$ is defined by (\ref{Tmax}), then the feedback of the produced particles on $\varphi$ is negligible, and the considerations from section \ref{SimplePerturbaiveSec} apply. In particular, the maximal temperature $T_{MAX}$ is given by $T_{max}$, and the reheating temperature is given by (\ref{TR}). 
\item If $T_R<T_1<T_{max}$, then the temperature $T_{max}$ predicted by (\ref{Tmax}) lies outside the range of applicability of that equation because the $\Gamma_2$-term in (\ref{GammaExpansionInn}) cannot be neglected.
For $x<x_1$, $R$ is given by (\ref{R0}). For later times, $R$ is given by \cite{Drewes:2014pfa}
\begin{eqnarray}
R&=&\theta(x_1-x) A_0\frac{2}{5}(x^{5/2}-1)
+ \theta(x-x_1)\left(A_2\left( x^{1/2} - x_1^{1/2} \right) + R_1^{1/2}\right)^{2}\label{R2},
\end{eqnarray}
which remains valid until $T$ drops below $T_1$ again. The reheating temperature is given by (\ref{TR}).
The maximal temperature $T_{MAX}$ can easily be determined from this expression and is given by 
\begin{eqnarray}
\tilde{T}_{max}&=&m_\phi x_1^{-3/4}\left(\frac{3}{4}\right)^2\left(\frac{A_2}{3}\right)^{1/2}\left(\frac{30}{\pi^2g_*}\right)^{1/4}\nonumber\\
&\simeq&0.33\sqrt{M_P\frac{\Gamma_2}{g_*}}\frac{\V_I^{1/4}}{m_\phi}\label{TtildeMax},
\end{eqnarray}
where $\V_I$ is the value of $\V(\varphi)$ at the end of inflation.
Interestingly, there is no huge effect on $\Rrad$ in spite of the significant change of the temperature evolution. The reason is that the duration of the reheating phase (and hence $\rhoreh$) is only mildly affected by this change as long as $T_R<T_1$ because the fraction of inflaton's energy loss is insignificant prior to the moment when $\Gamma_\varphi=H$, which occurs at $T=T_R$.
The change in $\wrehbar$ is also small because the total energy density is dominated by $\rho_\varphi$ before this moment.
\item Finally, if $T_1<T_R<T_{max}$, then the universe gets reheated very quickly due to a "thermal resonance", which happens in spite of the efficient rescatterings of produced $\chi$-particles.
In this case, the maximum temperature and reheating temperature are both roughly given by $\tilde{T}_{max}$ because the universe reheats almost instantaneously by the time $x=x_{\rm crit}$ 
\begin{equation}
x_{\rm crit}= \left[\frac{A_2\sqrt{x_1}-\sqrt{R_1}}{A_2-\sqrt{\Phi_I}}\right]^2\label{xcrit}
,\end{equation} 
with $\Phi_I=\V_I/m_\phi^4$ being the initial value of $\Phi$.
This happens if 
\begin{equation}
\Gamma_2 > \frac{m_\phi^2}{M_P}\left(\frac{8\pi^3g_*}{90}\right)^{1/2}\label{Gamma2crit}.
\end{equation}
In this case, there is an important effect on $\Rrad$ because the duration of the reheating process is considerably shortened, and $\rhoreh$ is larger than suggested by the estimate (\ref{TR}).
This very efficient reheating mechanism was first proposed in \cite{Drewes:2014pfa}. However, it remains to be seen if it can be realised without violating the adiabaticity condition (\ref{ADIABATICITY}), as large $\Gamma_2$ tend to require large couplings $\lambda$, which  tend to lead to violations of the adiabaticity condition due to contributions $\sim\lambda\varphi$ of the $\mathcal{X}$-masses. Moreover, it is questionable if the approximation of instant thermalisation is applicable during such a rapid reheating phase.
\end{itemize}
The key point is that the $\Gamma_2$-term comes from $\phi$-annihilation in inelastic scattering processes in the plasma and necessarily depends on $\alpha$ and other parameters in $\mathcal{L}_{\mathcal{X}}$.
Hence, also $\Gamma_\varphi$ depends on the parameters in $\mathcal{L}_{\mathcal{X}}$, and not only on those in $\mathcal{L}_\phi$. 
This dependence can only be neglected for $T_R<T_{max}<T_1$. Using the definitions of $T_{max}$ and $T_1$ and $\V_I\simeq m_\phi^2M_P^2/2$, this roughly corresponds to $\Gamma_2^2<11\Gamma_0m_\phi (m_\phi/M_P)^2 g_*$. Since $\Gamma_0$ and $\Gamma_2$ are both of second order in $\lambda$, we can (ignoring various prefactors) estimate that this holds for $\lambda^2\ll 11(m_\phi/M_P)^2g_*$, i.e.\ for rather small $\lambda$.
 
However, though the thermal history can be significantly modified for $T_{max}>T_1$, $\Rrad$ is only affected mildly unless $T_1<T_R$ because the modifications of the thermal history have no large effect on $\wrehbar$ as long as $\rho_\varphi\gg \rho_\gamma$. 
Assuming that $\Gamma_2\sim \lambda^2m_\phi$ and using (\ref{Gamma2crit}), the situation $T_1<T_R$, in which the thermal feedback changes the duration of reheating significantly, is only realised for $\lambda \gg m_\phi/M_P$.  
On the other hand, (\ref{adiabaticitysimple}) suggests that the adiabaticity condition only holds if $\lambda\ll (m_\chi/m_\phi)^3 m_\phi/M_P\ll  m_\phi/M_P$, where we have again estimated $\varphi\sim M_P$ at the beginning of reheating and assumed $m_\chi<m_\phi$. 

These very simple estimates suggest that $\Rrad$ is only affected significantly by the parameters in $\mathcal{L}_{\mathcal{X}}$ if $\lambda$ is large enough that the particle production enters a resonance.
If $\lambda$ is large enough that simple estimates along the lines of (\ref{NarrowResonanceGamma}) suggest that there may be a resonance, then $\Rrad$ becomes sensitive to the parameters in $\mathcal{L}_{\mathcal{X}}$ because these determine whether scatterings amongst the produced particles can prevent or delay the resonance.
If, on the other hand, $\lambda$ is small, then there is no resonance.  
Even though the conditions $i)$ and $ii)$ can both also be violated during perturbative reheating, the effect of the parameters in $\mathcal{L}_{\mathcal{X}}$ on $\Rrad$ is generally negligible.

\subsection{Phase space blockings}\label{PhaseSpaceBlockingsSec} 
The dispersion relations $\Omega_{\mathcal{X}}$ of quasiparticles in a plasma are not identical to the on-shell energy $\omega_{\mathcal{X}}^2=\textbf{p}^2+m_{\mathcal{X}}^2$  of particles with mass $m_{\mathcal{X}}$ in vacuum. 
In addition to the time-dependent contribution to  $\Omega_{\mathcal{X}}$ from the interactions with the background field shown in (\ref{OmegachiDef}), there are also thermal corrections from forward scatterings in the plasma. 
This provides an additional source of non-trivial $T$- and $\alpha$-dependence of $\Gamma_\varphi$. 
In the simplest case, the effect of the medium can be described by a $T$- and $\alpha$-dependent, but momentum independent thermal mass, such as (\ref{ThermalMass}). 
In general, in particular in gauge theories, the momentum dependence is more complicated \cite{Klimov:1982bv,Weldon:1983jn,Thoma:1994yw,Drewes:2013bfa}, and there are new collective excitations in addition to the screened one-particle states.
The thermal masses enter all calculations at finite temperature.

The most prominent effect occurs if reheating is mainly driven by $1\rightarrow 2$ decays and $\alpha$ is relatively large. Then sum of the decay products' thermal masses scales as $\propto \alpha T$ and can exceed the effective inflaton mass (which grows slower due to the weaker interactions of $\phi$) at some critical temperature $T_c$. This blocks the decay kinematically, making reheating inefficient for $T>T_c$. 
Hence, the temperature in the early universe cannot significantly exceed $T_c$  if reheating is driven by perturbative decays, regardless of $\V_I$. 
Of course, this is not exactly true because there may be other channels than the $1\rightarrow 2$ decay, but there exists scenarios in which the phase space blocking can be efficient. This effect was first pointed out in \cite{Kolb:2003ke} and further investigated in \cite{Yokoyama:2005dv,Drewes:2010pf}. Finally, a detailed analysis in \cite{Drewes:2013iaa} concluded that this effect can indeed be relevant during reheating and, for instance, occurs in the toy model (\ref{toymodel}) with (\ref{Lchi}) and $\alpha\gg\lambda$.
For $T_c<T_R$ the phase space modification has no big effect on $\Rrad$  because $\rho_\varphi\gg\rho_\gamma$ until $\Gamma_\varphi=H$, and even huge changes in the $T$-evolution cause only a mild change in $\wrehbar$. For $T_c<T_R$, on the other hand, the universe cannot reach the would-be reheating temperature, and the reheating phase can be prolonged, leading to a smaller $\rhoreh$ that affects $\Rrad$.
Since the thermal masses depend on $\alpha$ and $T$, they lead to a dependence of $\Gamma_\varphi$ on these parameters and can violate conditions $i)$ and $ii)$.

The situation $T_c<T_R$ can only be realised for relatively large $\lambda$ and $\alpha$. Ignoring numerical prefactors that depend on the details of the interactions, we can estimate the thermal masses as $M_{\mathcal{X}}\sim\alpha T$, which is e.g. seen in gauge theories \cite{Weldon:1982bn,LeB}.
If we furthermore assume that we can use the vacuum mass for $\phi$ (because the hierarchy $\lambda\ll \alpha<4\pi$ keeps thermal corrections small), one can estimate $T_c\sim m_\phi/(2\alpha)$. Comparing this to $T_R$, we find the condition 
\begin{equation}
\Gamma_0\alpha^2>\frac{m_\phi^2}{4 M_P}\sqrt{\frac{8\pi^3 g_*}{90}}
\end{equation}
Since usually $\Gamma_0\propto \lambda^2 m_\phi$, this requires $\alpha^2\lambda^2\gg m_\phi/M_P$. We can again compare this to the condition (\ref{adiabaticitysimple}), which suggests the adiabaticity condition only holds if $\lambda\ll (m_\chi/m_\phi)^3 m_\phi/M_P$. Using $m_\chi<m_\phi$, it seems that phase space blockings do not affect $\Rrad$ strongly in scenarios where reheating is perturbative.

\subsection{Parametric resonance and combined reheating}\label{CombinedReheatingSec}
So far we have assumed that $\varphi$ dissipates its energy via perturbative processes. It is, however, well-known that particles can also be produced nonperturbatively if the adiabaticity condition (\ref{ADIABATICITY}) is violated. From the example (\ref{OmegachiDef}) it is clear that this generally happens if the coupling constants $\lambda$ are larger than some critical value, which depends on the terms in $\mathcal{L}_{\rm int}$. 
The effect can easily be described in the model (\ref{toymodel}).
The coupling to $\varphi$ in (\ref{OmegachiDef}) gives a time dependent tree level mass to $\chi$. Neglecting Hubble expansion, the Klein-Gordon equation for the mode functions reads 
\begin{equation}
\ddot{\chi}_k(t) + \left[\textbf{k}^2 + m_\chi^2 
+ m_\phi\lambda_3\varphi(t)\right]\chi_k(t)=0
\end{equation}
This can be rewritten in terms of the variable $z=m_\phi t/2$ as 
\begin{equation}
\chi_k''(z) + \left[
A_k - 2q\cos(2z)
\right]\chi_k(z)=0\label{Mathieu}
\end{equation}
with
\begin{eqnarray}
A_k=\frac{4\omega_k^2}{m_\phi^2} \ , \ q=-2\lambda_3\frac{\varphi|_{z=0}}{m_\phi}
\end{eqnarray}
This is the well-known Mathieu equation. It shows instabilities that can be characterised by the values of $A_k$ and $q$. 
The exponentially growing solution can in interpreted as resonant particle production. A quantitative treatment requires quantum field theory in a time dependent background, but for the present purpose it is sufficient to qualitatively distinguish the textbook cases $|q|\ll 1$ and $|q|>1$. In the former case, resonant particle production only occurs for specific momentum modes, similar to the perturbatively enhanced rate (\ref{NarrowResonanceGamma}).
This effect tends to be suppressed by Hubble expansion, which redshifts the produced particles away from these modes and thereby reduces the Bose enhancement that feeds the resonance. A similar effect is achieved by scatterings amongst the $\chi$-particles.
For $|q|>1$ the adiabaticity condition is violated for all modes below a value 
$k_*$, which can be estimated as $k_*^2\sim (\lambda_3 \varphi m_\phi^2)^{2/3}-m_\chi^2$ by using (\ref{adiabaticitysimple}). That is, particle are produced efficiently in all modes with $|k|<k_*$. In contract to the narrow resonance, Hubble expansion in this case helps to maintain the \emph{broad resonance} because it redshifts $\chi$-particles that have gained momenta $>k_*$ back into the sphere $|\textbf{k}|<k_*$.
This parametric resonance leads to very efficient particle production that can dissipate most of $\rho_\varphi$ into radiation in a few oscillations.
It strongly violates condition $i)$, as $\Gamma_\varphi$ is strongly affected by the (time dependent) feedback of the produced particles. 
However, since $A_k$ and $q$ only depend on $\V(\varphi)$ and $\lambda$ (and not on $\alpha$), one might hope that at least condition $ii)$ holds and the resonance does not depend on $\alpha$ and the particle content of $\mathcal{L}_{\rm full}$.
This would allow to translate a bound on $\Rrad$ into a bound on $\lambda$, though the relation can only be found numerically.

However, the produced $\chi$-particles undergo inelastic scatterings or can decay into secondary particles.
Let us assume that there is another field $\psi$ that does not directly couple to the inflaton, but interacts with $\chi$ via some coupling $\alpha$.
Nonperturbative particle production usually occurs around the zero-crossings of $\varphi$, when the adiabaticity condition is violated. 
For larger elongations of $\varphi$, where $\dot{\varphi}$ is smaller and the oscillating field spends most of its time, the system is adiabatic and all processes can be described perturbatively. Large elongations of $\varphi$ make the primary decay products $\chi$ heavy because of their direct coupling to $\varphi$. 
Their effective mass in (\ref{OmegachiDef}) receives a contribution $\sim\lambda\varphi$. 
The secondary decay products tend to have smaller effective masses because they do not directly couple to $\varphi$.
If $\chi$-particles can decay into $\psi$-particles before the next zero crossing, the Bose enhancement that drives the parametric resonance is reduced.
We may estimate that the lifetime of the primary decay products scales as $\tau_{\mathcal{X}}=\Gamma_{\mathcal{X}}^{-1}\sim \alpha^{-2}M_{\mathcal{X}}^{-1}\sim \alpha^{-2}\lambda^{-1}\varphi^{-1}$, where we have ignored any thermal corrections.
That is, for large $\varphi$-elongations, the $\chi$ particles become very short lived, and most of them may decay before the next zero-crossing of $\varphi$.
In \cite{GarciaBellido:2008ab} it has been shown that this effect can delay or even completely suppress the parametric resonance. Obviously, this depends on $\alpha$ and the properties of the $\psi$ particles. A similar effect is achieved by inelastic scatterings, which have not been considered in \cite{GarciaBellido:2008ab} and related studies \cite{GarciaBellido:2012zu,Enomoto:2014cna}. 
This implies that the nonperturbative particle production depends on the particle physics model into which $\V(\varphi)$ and $\mathcal{L}_{\rm int}$ are embedded.
Since this dependence arises due to an interplay of perturbative effects (particle decays and scatterings) and non-perturbative preheating, it has been dubbed \emph{combined reheating}.

\section{Discussion}\label{SecDiscussion}
Let us come back to the questions 1)-4) posed in section \ref{Nepenthe}.

\paragraph{1) If one specifies a cosmological model  of inflation, which role does the reheating era play when extracting information about $\V(\varphi)$ during inflation from the CMB?} -
If only $\V(\varphi)$ is specified, then it is not possible to reverse-engineer $\mathcal{L}_{\rm full}$ in a unique way, which would be required to calculate $\Gamma_\varphi$ and $\Rrad$. 
This does not only make it difficult to learn something about the shape of $\V(\varphi)$ near its minimum from reheating, it also imposes a fundamental limitation (or systematic uncertainty) to the  extraction of information on the effective potential $\V(\varphi)$ during inflation from the CMB. 
The reason is that (\ref{preRradDef}) requires knowledge of $\Rrad$ in order to translate observed CMB anisotropy multipoles into physical scales during inflation.
This should be seen as a systematic uncertainty that cannot be eliminated without specifying a particle physics model $\mathcal{L}_{\rm full}$.

Moreover, the derivation of (\ref{preRradDef}) shows that $\Rrad$ can only be uniquely related to reheating after inflation
if no significant amounts of entropy were injected into the primordial plasma between that moment and the CMB decoupling (apart from the SM processes that we know how to describe). 
There are many particle physics models in which this assumption is not fulfilled because some heavy particles or moduli decay out of equilibrium during this time.

\paragraph{2) If one specifies a cosmological model of inflation, what can the CMB tell about the reheating phase?} -
As discussed in 1), our ignorance about reheating imposes a systematic uncertainty when constraining inflationary models from CMB data, which cannot be eliminated without knowledge of $\mathcal{L}_{\rm full}$.
One may wonder whether it is possible to turn the tables and learn something about the reheating era from the CMB for a given model $\V(\varphi)$, without knowledge of $\mathcal{L}_{\rm full}$.
In \cite{Martin:2010kz} it has been shown that one can derive the non-trivial bound (\ref{Trconstraint},\ref{Trconstraintp}) on the reheating temperature from the ``trivial'' requirement (\ref{boundaryw1})-(\ref{boundarynuc}) that reheating happens after inflation and before BBN.
In the same work, it has also been suggested that $\Rrad$ could, rather than being calculated, be treated as a free fit parameter in a Bayesian analysis.
The choice of prior for $\Rrad$ is not completely arbitrary because it has to obey the boundary conditions imposed by Eqns.~(\ref{boundaryw1})-(\ref{boundarynuc}), though that still leaves some freedom of how to extrapolate between them (e.g. flat on logarithmic or in linear scales). The philosophy of \cite{Martin:2010kz} is to consider it as "evidence" for certain values of $\Rrad$ if the posterior is more peaked than the prior. The interpretation is then that an inflationary model characterised by $\V(\varphi)$ with a reheating history characterised by $\Rrad$ gives a better fit to the data than the same $\V(\varphi)$ with a different expansion history during reheating.
In this sense, $\Rrad$ should simply be understood as an additional parameter in all inflationary models. 
In simple large field models, $\wrehbar$ is in good approximation fixed once $\V(\varphi)$ is specified, and a constraint on $\Rrad$ is essentially a constraint on $\rhoreh$ and therefore $T_R$.

\paragraph{3) How can the imprint of reheating in the CMB help to determine which of the known models (if any) is realised in nature?} -
It is instructive to pose the question:
Are there inflationary models $\V(\varphi)$ that appear to be consistent with CMB data if one ignores the effect of reheating, but are excluded if constraints from reheating are taken into account?\footnote{
It is obvious that reheating considerations can exclude models my means other than the CMB if observable relics are produced during reheating.
This is e.g. the case if a certain model relies on reheating in order to produce the observed Dark Matter density, or if the baryon asymmetry of the universe is to be generated during reheating, but a detailed calculation shows that the model fails to do so.
Similarly, a model is obviously excluded if it predicts relics (heavy particles, topological defects, gravitational waves etc.) that are not observed, see e.g. \cite{Amin:2014eta} for a recent review.}
That is, can models appear viable if $\Rrad$ is taken to e unity, but are excluded based on CMB data by using realistically accessible information about $\Rrad$? 
Indeed, if the window (\ref{Trconstraint})-(\ref{Trconstraintp}) of allowed values for $\Rrad$ does not overlap with the range allowed by (\ref{boundaryw1})-(\ref{boundarynuc}), then a model $\V(\varphi)$ can  be excluded. 
Again assuming that the relation between $V(\phi)$ and $\V(\varphi)$ is known, this can also be used to exclude some classes of microphysical models.
It seems, however, unlikely that many of the popular choices for $\V(\varphi)$ can be excluded without additional input from microphysics. The previously discussed dependencies of $\Rrad$ on the parameters in $\mathcal{L}_{\mathcal{X}}$ imply that the range of $\Rrad$ probably cannot be predicted much more precisely than (\ref{boundaryw1})-(\ref{boundarynuc}) without detailed calculations in specific particle physics models $\mathcal{L}_{\rm full}$.

If one wants to be less ambitious than exclusion, one could ask whether $\Rrad$ can be used to say anything about the likelihood of any model to be correct. As always, any statements about the measure in parameter space or the probability that one or another model is correct are highly subjective. 
One way to turn this question into a quantitative statement is offered by Bayesian statistics. In \cite{Martin:2014nya} the idea to treat $\Rrad$ as a free fit parameter in a Bayesian approach has been applied to a range of models listed in \cite{Martin:2013tda}. Rather than just looking for a change in the posterior vs. prior   of $\Rrad$ in order to learn something about $T_R$ in a particular model, one can use the constraints (\ref{boundaryw1})-(\ref{boundarynuc}) on $\Rrad$ as an additional piece of information that affects the overall compatibility of one model with data with respect to another model. This can be interpreted as a measure for the likelihood that one model is correct (compared to another).

The analysis in \cite{Martin:2014nya} shows that the ``trivial'' requirements (\ref{boundaryw1})-(\ref{boundarynuc}) alone lead to improved predictions for parameters. 
However, though there are significant constraints on reheating in some specific classes of models (such as loop inflation), it does not allow to draw any strong conclusions regarding the correct model of inflation. 
The reason for the relatively small improvement from including $\Rrad$ is that the equations (\ref{boundaryw1})-(\ref{boundarynuc}), which contain the only reheating-related piece of information that entered the analysis, 
are not sensitive to any details of the reheating process or the underlying model. 
The authors of that work point out that, if the understanding of reheating improves, one may be able to impose stronger boundary conditions on $\Rrad$ that make a bigger difference.
However, as pointed out before, it is in general impossible to calculate $\Gamma_\varphi$ from the knowledge of $\V(\varphi)$ and $\mathcal{L}_\phi$ alone. 
Therefore it seems likely that stronger boundary conditions on $\Rrad$ can only be obtained if $\mathcal{L}_{\rm full}$ is specified. If this is done, then the Bayesian analysis in fact is a comparison of particle physics models $\mathcal{L}_{\rm full}$ rather than inflationary models $\V(\varphi)$.

\paragraph{4) If one specifies a particle physics model, can the information on the reheating phase that the CMB contains be used to improve constraints on model parameters?} -
If one specifies a complete particle physics model $\mathcal{L}_{\rm full}$, then it is at least in principle possible to solve the nonequilibrium problem of reheating 
and calculate both, $\V(\varphi)$ and  $\Rrad$. 
If one can express $\Rrad$ in terms of the parameters in $\mathcal{L}_{\rm full}$ (in particular $\lambda$, $\alpha$ and the parameters in $V(\phi)$), 
then this relation can not only be used to overcome the systematic uncertainty in extracting information about $\V(\varphi)$ during inflation from the CMB, but also to impose bounds on the model parameters from reheating. The constraints on individual parameters may be very weak because the single observable $\Rrad$ can depend on a comparable large number of parameters, leading to degeneracies. 
The possibility to recover any information about physics during the reheating stage would nevertheless be very interesting.

Practically it can be very difficult to determine $\Rrad$ and $\V(\varphi)$.
If there is a parametric resonance or other instability, it is only possible with extensive numerical calculations. 
Moreover, a conceptual problem is that most particle physics theories contain many other degrees of freedom in addition to $\phi$ and the SM, and it is not uniquely fixed how these couple to each other, $\phi$ and the SM.\footnote{Higgs inflation is one of the few exceptions for which all interactions are know, and that is the reason why reheating in this scenario has been studied in some detail \cite{GarciaBellido:2008ab}.}
In numerical calculations, specification of $\mathcal{L}_{\rm full}$ does not only require to write down the complete Lagrangian, but also to fix the numerical values of all parameters.
The problem is that the reheating process is rather sensitive to the parameters $\lambda$ and $\alpha$ even at the qualitative level. They e.g. determine whether or not a parametric resonance occurs. Within the very same model $\mathcal{L}_{\rm full}$, a change in the numerical value of the $\lambda$ can lead to a completely different reheating.
Given the large number of models, particle physicists are usually not interested in calculating all details of a specific model for many different parameter choices. A more relevant question is whether successful inflation and reheating can at all be accommodated in a given class of particle physics theories (Is inflation possible in minimal supergravity? Can viable axion monodromy inflation be embedded in string theory?).

While these considerations make it unlikely that the CMB can be used to understand the microphysics of a preheating phase in the foreseeable time, the situation seems much more encouraging in scenarios where reheating is entirely driven by perturbative processes. 
As the examples in sections \ref{SimplePerturbaiveSec}-\ref{CombinedReheatingSec} show, the dependence of $\Rrad$ on the parameters in $\mathcal{L}_{\mathcal{X}}$ tends to be negligible if there is no parametric resonance or other instability. Then $\Rrad$ is essentially given by $\lambda$ and the parameters in $V(\phi)$ alone. Of course, there are still parameter degeneracies, and it is obvious that even an exact measurement of one single parameter $\Rrad$ would not allow to fix several inflaton couplings. 
However, the number of microphysical parameters upon which $\Rrad$ depends is comparably small, and this dependency may be understood analytically (or at least with considerably less numerical effort than in the case of preheating).
This means that it can be possible to extract meaningful constraints on the parameter space of microphysical models $\mathcal{L}_\phi$ from the CMB. For example, in the simple model discussed in section \ref{SimplePerturbaiveSec}, an analytic relation (\ref{RexplicitResult}) between $\Rrad$ and $\lambda$ could be established.\footnote{Our discussion has been focused on scenarios in which reheating occurs during damped oscillations near the minimum of $\V(\varphi)$. There are several other possibilities to heat up the universe. For instance, in string theory reheating could occur more or less instantaneously through the annihilation of two branes. 
Another possibility is that inflation ends with a ``tachyonic'' field configuration, and the universe is reheated while this configuration relaxes to the true vacuum, see e.g. \cite{GarciaBellido:1999sv,Felder:2000hj}. 
If the transition to the true vacuum happens non-adiabatically, then reheating may be driven by a via tachyonic (or spinoidal) instability. Similar to the parametric resonance, strong feedback effects and nonperturbative physics probably make it very difficult to establish a relation between $\Rrad$ and microphysical parameters. If the transition to the true vacuum happens adiabatically and reheating is driven by perturbative physics, then one may determine such a relation, and the CMB might be used to impose constraints on microphysical parameters.
}
This allows to constrain the inflaton coupling $\lambda_3$ using knowledge of $\Rrad$ obtained from CMB observations.
Due to the relation (\ref{TRestimate}), this knowledge can also be used to constrain the reheating temperature.

\section{Conclusions}\label{Conclusions}

The reheating phase affects the CMB via its effect on the expansion history of the universe.
This effect can be parametrised in terms of the parameter $\Rrad$. 
The trivial requirements that reheating occurs after inflation and before BBN allow to use $\Rrad$ to impose a non-trivial constraint on the reheating temperature for given effective inflaton potential $\V(\varphi)$: The range of values for $\Rrad$ that is consistent with these requirements must overlap with the range of values for which the predictions of $\V(\varphi)$ for CMB anisotropies are consistent with data. If the overlap between these two windows is smaller than the former, then this implies a non-trivial bound on the reheating temperature $T_R$.

To go beyond that, many phenomenological studies of reheating adopt the assumptions that the inflaton dissipation rate $\Gamma_\varphi$ is $i)$  constant in time during reheating and $ii)$ independent of the properties of the primordial plasma. 
Assumption $ii)$ implies that also $\Rrad$ depends only on the inflaton couplings,  i.e.\, the parameters in $\mathcal{L}_\phi$.
In this case a constraint on $\Rrad$ can directly be converted into a constraint on these parameters. Vice versa, if $\Rrad$ is calculable, then this allows to eliminate the uncertainty that the reheating phase brings into the extraction of information about $\V(\varphi)$ during inflation from the CMB.

In the previous sections, we have illustrated that the assumptions $i)$ and $ii)$ are not valid in general. They can both be violated in several ways by the feedback of the produced radiation on the evolution of the inflaton field $\varphi$, which leads to the different phenomena discussed in sections \ref{NarrowResonanceSec}, \ref{PhaseSpaceBlockingsSec} and \ref{CombinedReheatingSec}. 
This can significantly affect the thermal history during reheating.
While this seems unsurprising in scenarios where reheating is driven by nonperturbative far-from-equilibrium dynamics, we showed in sections \ref{NarrowResonanceSec} and \ref{PhaseSpaceBlockingsSec} that assumptions $i)$ and $ii)$ can also be violated in scenarios where reheating is driven my simple perturbative processes, such as inflaton particle decays.
The dependence of $\Gamma_\varphi$ on the properties of the primordial plasma (i.e.\ the microphysical parameters of the particle physics model $\mathcal{L}_{\mathcal{X}}$) is carried over to $\Rrad$ and can therefore affect CMB observables.  
On one hand, this means that it may be impossible to eliminate the uncertainty that the reheating phase imposes on CMB-measurements of $\V(\varphi)$ without specifying the details of the particle physics model into which $\V(\varphi)$ is embedded. 
On the other hand, it also means that it is difficult to extract any information about the reheating phase itself from the CMB.

However, this does not mean that all hope is lost. 
The reason is that $\Rrad$ is only sensitive to the \emph{expansion history}. 
If the total energy density during reheating is dominated by the inflaton's energy $\rho_\varphi$, even large changes in the \emph{thermal history} (temperature evolution and the equation of state of the produced radiation) leave $\Rrad$ more or less unaffected as long as the reheating temperature (or $\rhoreh$) is not changed. 
It is therefore important to distinguish how the violation of $i)$ and $ii)$ affects the thermal and expansion history.
In those models where the inflaton couplings $\lambda$ are large enough to trigger a resonant production of particles, the expansion history and reheating temperature are indeed sensitive to the properties of the primordial plasma. Then $\Rrad$ depends on a large number of microphysical parameters in  $\mathcal{L}_{\mathcal{X}}$ in a complicated way. This makes it unlikely that much more information beyond what has been discussed here can be extracted from the CMB without detailed, computationally very expensive numerical studies for different parameter choices in each particle physics model. 

If, on the other hand, there is no parametric resonance and reheating happens via perturbative processes, then our results imply that the CMB can be used to constrain the inflaton couplings and the reheating temperature.
Even though the feedback of the produced particles might affect the temperature evolution in the early phase of reheating, $\rhoreh$ is rather insensitive to the parameters in $\mathcal{L}_{\mathcal{X}}$.
This means that $\Rrad$ is calculable and fixed by the parameters in the inflaton Lagrangian $\mathcal{L}_\phi$ alone, and constraints on $\Rrad$ can directly be translated into constraints on the inflaton couplings. The CMB can therefore provide an indirect probe of fundamental microphysical parameters that most likely can never be measured directly in the laboratory, but have an immense impact on the evolution of the cosmos by setting the stage for the hot big bang.
The improved understanding of the reheating phase also allows to reduce the systematic uncertainty due to reheating when extracting constraints on $\V(\varphi)$ during inflation from the CMB. Finally, the knowledge about the inflaton couplings can be used to constrain the reheating temperature.
Though this conclusion is only based on a few simple examples of large field models, the parametric dependencies suggest that it might hold for most single field models in which reheating is a perturbative process.

\section*{Acknowledgements} 
We would like to thank Christophe Ringeval, Jerome Martin, Vincent Vennin, Jan Hamann and Eiichiro Komatsu for helpful comments and discussions.
This work was supported by the Gottfried Wilhelm Leibniz program of the Deutsche Forschungsgemeinschaft (DFG), the DFG cluster of excellence Origin and Structure of the Universe, the Projektbezogener Wissenschaftleraustausch program of the Bayerisches Hochschulzentrum f\"ur China and the visitor program of the Kavli Institute for Theoretical Physics China (KITPC) in Beijing.

\bibliographystyle{JHEP}
\bibliography{all}

\end{document}